\documentclass[12pt,a4paper,aps,prb,superscriptaddress]{revtex4}

\usepackage{epsfig}

\begin{document}


\thispagestyle{empty}

\title{Band Gap of CsCl Film's as a Function of Lattice Constant.} 
\author{Kuldeep Kumar, \& P.Arun}
\affiliation{Department of Physics \& Electronics, S.G.T.B. Khalsa College, 
University of Delhi, Delhi-110 007, India}
 \email{arunp92@physics.du.ac.in}
\author{N.C.Mehra}
\affiliation{University Science \& Instrumentation Center, 
University of Delhi, Delhi-110 007, India}
\author{L.Makinistian}
\affiliation{Facultad de Ingenieria,
Universidad Nacional de Entre Rios,
3101 Oro Verde (ER),
Argentina}
\affiliation{Grupo de Materiales Computacionales,
INTEC-CONICET,
Guemes 3450-3000 Sante Fe,
Argentina}
\author{E.A.Albanesi}
\affiliation{Facultad de Ingenieria,
Universidad Nacional de Entre Rios,
3101 Oro Verde (ER),
Argentina}
\affiliation{Grupo de Materiales Computacionales,
INTEC-CONICET,
Guemes 3450-3000 Sante Fe,
Argentina}

\begin{abstract}
\vskip 1cm
 The manuscript reports
experimental estimation of Cesium Chloride film's band gap and compares it 
with theoretical simulations. The band gap shows a strong dependence on it's
lattice constant. Also, it is seen that the point defects present in the 
films strongly affects it's optical properties.
\end{abstract}

\maketitle


\section{Introduction}

In the past Alkali Halides (AH) were a topic of interest for crystallographers and
researchers involved in energy band structure calculations. However,
recently interest in AH have been renewed in connection with their optical
properties for device applications\cite{r8} and optical wave
guides.\cite{r9} Due to solubility of
CsCl in its thin film state, Yoshikawa $et al$\cite{r77} did some limited 
amount of in situ characterization. Tsuchiya $et al$\cite{r17} have however 
taken advantage of CsCl's high
solubility and used its thin films as steam etch-able resists in IC fabrication.
With such new applications of AH and in particular CsCl thin films and the 
lack of any extensive
study of these materials in its thin film state, we were encouraged to study 
the optical properties of Cesium Chloride in thin film 
state for the present work. The present work consist details of structural, 
morphological and optical spectrum analysis done on Cesium Chloride films. 

\section{Experimental}
Thin films of Cesium Chloride were grown by thermal evaporation 
method using 99.98\% pure starting material obtained from Loba Chemie Pvt Ltd, 
Mumbai. The films were grown on glass substrates kept at a distance 
of 12-15cm above the boat. The substrates were maintained at room 
temperature. Thin films were grown
at vacuum better than ${\rm 10^{-6}}$-${\rm10^{-7}}$Torr. Films grown by this 
method were found to be slightly translucent with bluish to
bluish-green tinge. The prelevance of color in these films were explained due 
to vacancies present in the films.\cite{rab,ava} Such vacancies would have
resulted in the films as a result of possible dissociation
of alkaline halide molecules when they are heated for evaporation.

\section{Results and Discussion}

\begin{figure}[h]
\begin{center}   
\epsfig{file=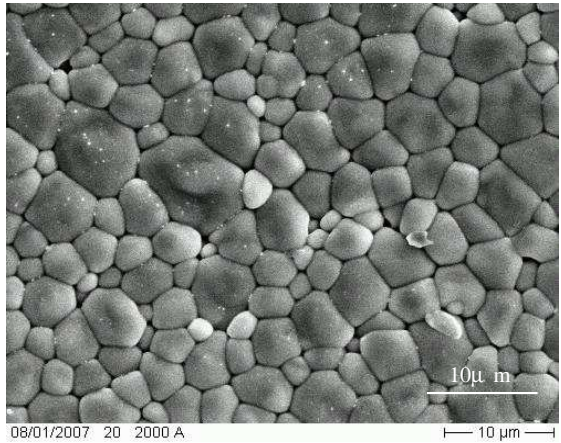,width=3in}
\vskip -.7cm
\caption{\sl The SEM micrograph of CsCl thin film.
It can be observed that the grains of various sizes are densely packed
throughout the film's surface.}
\vskip -1cm
\end{center}
\label{fig:sem}
\vskip -0.5cm  
\end{figure}   

The structural studies of our films were done using Philips PW 3020 X-ray 
diffractometer. The as grown films of CsCl were found to be polycrystalline 
in nature. The peak positions match those listed 
in ASTM Card No.05-0607. The polycrystalline nature of the films were
also confirmed by studying the surface morphology of the films (fig~1) using
JEOL (JSM)-840 Scanning Electron Microscope (SEM). 
Optical characterization of our samples were carried out using the 
UV-visible spectroscope (Shimadzu's UV 2501-PC).
Since CsCl has a very large band-gap, the absorption edge lies in the UV 
region. The absorption data in visible range, hence can not be used to 
determine the energy band gap using the conventional Tauc's method\cite{r218}.
However, indirect estimation of the band-gap can still be made from the
absorption spectra in a region away from the band edge. In fact, it has been 
shown that the optical
absorption coefficient of semiconductors and insulators vary exponentially
with incident photon energy away from the band edge.\cite{Urb} The behavior
is called the Urbach tail and the absorption coefficient is related to the
photon energy as\cite{kurik}
\begin{eqnarray}
\alpha(E)=\alpha_oexp^{\left({E_g-E \over E_u}\right)}\label{Urbach}
\end{eqnarray}
\begin{figure}[h]
\begin{center}
\epsfig{file=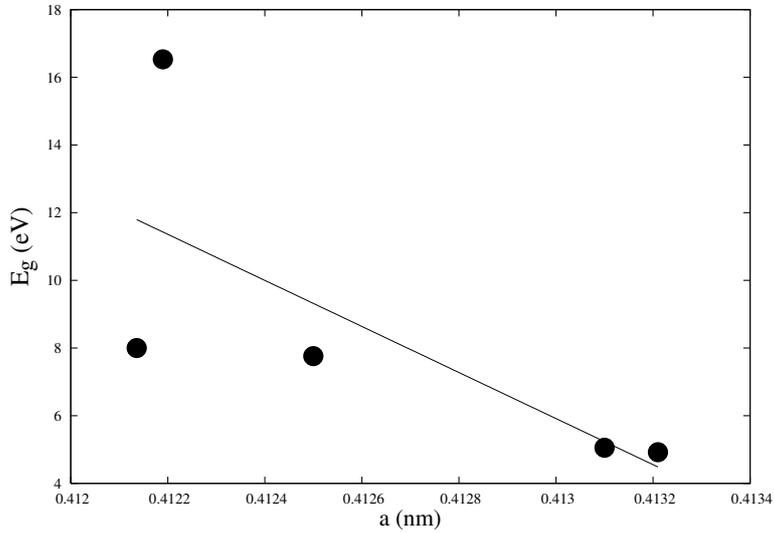,width=4in}
\vskip -.3cm
\caption{\sl The (estimated) band-gap of CsCl decreases with increasing
lattice constant.}
\vskip -1cm
\end{center}
\label{fig:7eg}
\vskip -0.5cm
\end{figure}

where ${\rm E_g}$ is the extrapolated band-gap and ${\rm \alpha_o}$,
the absorption coefficient of the material at the band-gap energy. ${\rm
E_u}$ is called the Urbach energy. 
The values of Urbach energy can be determined from the slope of the best fit 
straight line
\begin{eqnarray}
ln(\alpha)=\left[ln(\alpha_o)-{E_g\over E_u}\right]+\left({1\over
E_u}\right) 
E\label{Urbacha}
\end{eqnarray}
We have fit eqn(\ref{Urbacha}) on our absorption data and have estimated the 
values of ${\rm E_u}$. Using this result, the band gap of the films were 
extrapolated from the intercepts obtained by
fitting eqn(\ref{Urbacha}) to our data. A plot of such an estimated 
${\rm E_g}$ with
increasing lattice constants shows a decreasing trend (fig~2). 
The displacement of Cs and Cl atoms, moving away from each 
other with increasing lattice constant maybe the cause of decreasing band gap. 
Increasing atomic distances would decrease the effective lattice
potentials as seen by the electrons producing appreciable change in band gap.
\begin{figure}[h]
\begin{center}
\epsfig{file=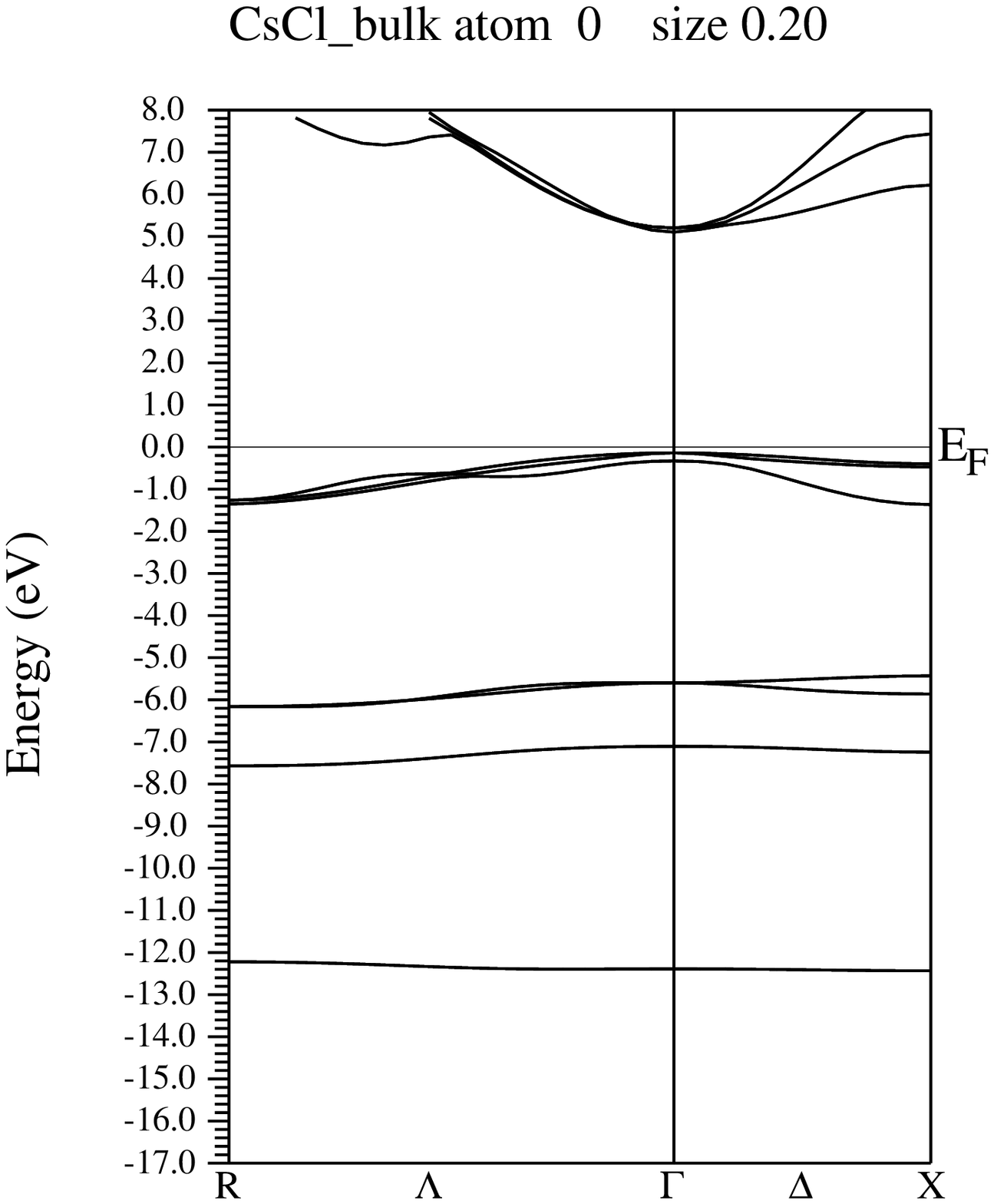,width=3in}
\epsfig{file=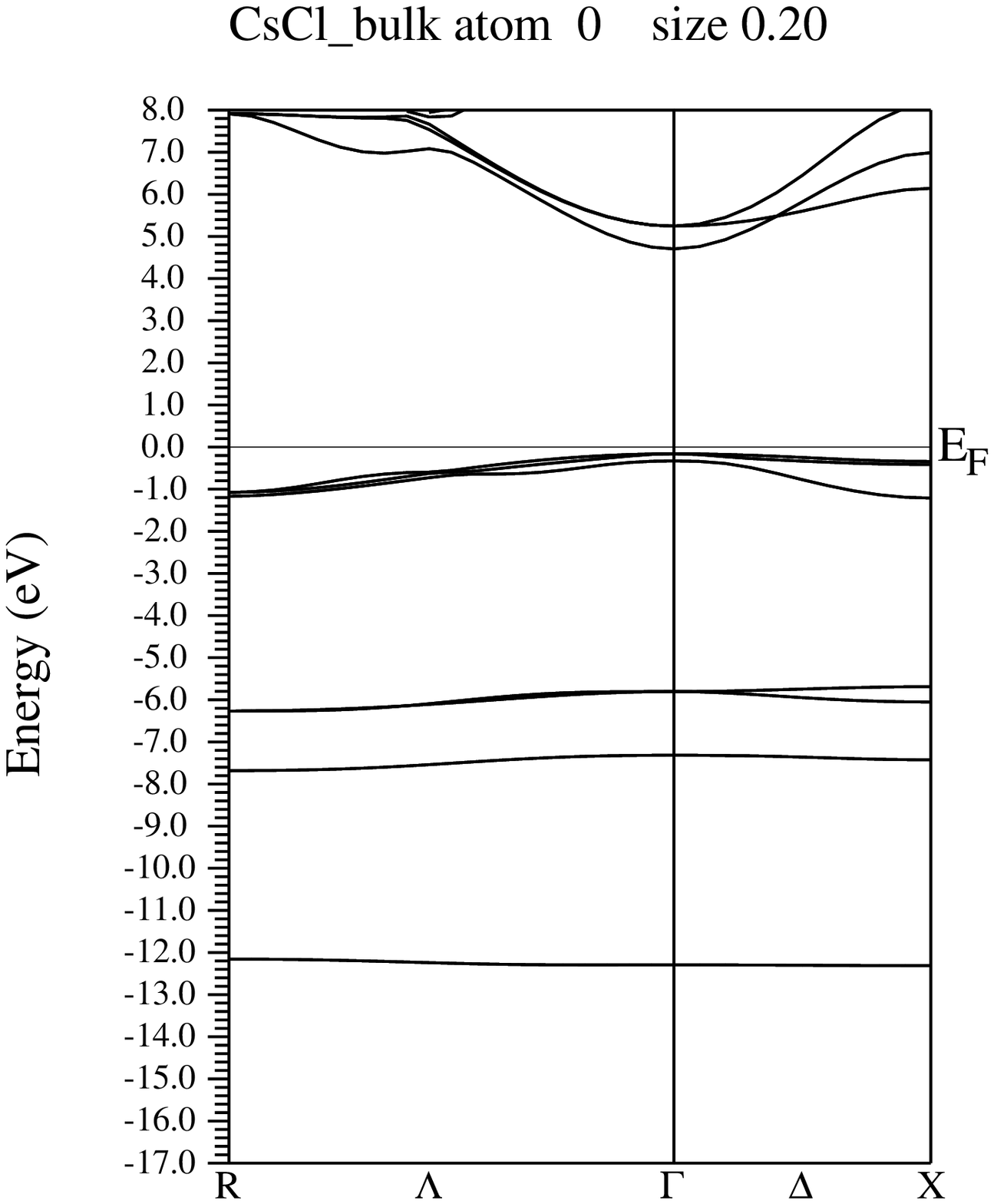,width=3in}
\vskip -1.4cm
\caption{\sl Calculated band structures at (a) lattice
constant of single crystal (0.4120nm) and (b) of lattice constant 0.4236nm.}
\vskip -1cm
\end{center}
\label{fig:struct}
\vskip -0.5cm
\end{figure}

To substantiate this we have theoretically calculated the band structure to 
investigate the
variation in energy band gap with varying lattice constant. 
For our calculations we have used the Full-Potential Linearized
Augmented Plane Wave (FP-LAPW) method within the 
density functional theory (DFT) as implemented by the WIEN2k code.
\cite{WIEN2k} 
In fig~3, we compare the band structures of bulk CsCl with lattice parameters 
(a) of CsCl in single crystal state (=0.412nm) and (b) at 0.4236nm. 
The figure shows the region R $\rightarrow$ ${\rm \Gamma}$
 $\rightarrow$ X that forms the minimum
gap. In both cases we have a direct transition band gap, located at the 
${\rm \Gamma}$ point.
The main difference corresponds to the bottom of the conduction band at
the ${\rm \Gamma}$ point, since at the equilibrium lattice constant three 
bands forming the
bottom of the conduction band are degenerated at the ${\rm \Gamma}$ point, 
while
at the higher lattice constant the degeneration is broken and a band
splits down in energy, with the main difference of 0.37 eV at the ${\rm
\Gamma}$ point. The breaking of degeneracy is observed in all simulations 
where the lattice constant is considered more than 0.412nm. The variation in
band gap with increasing lattice constant was theoretically simulated (see 
fig~4). The valence bands maintains its structure in all cases,
although negligible differences can be observed for the top of the
valence band in the ${\rm \Delta}$ direction.

\begin{figure}[h]
\begin{center}
\epsfig{file=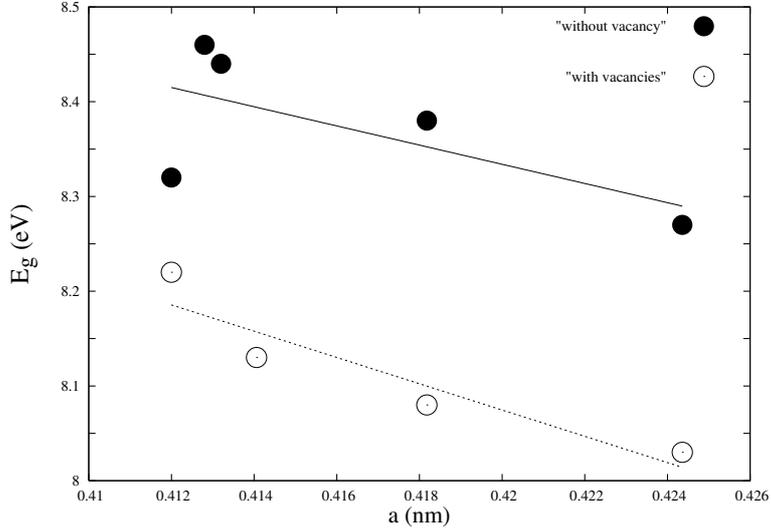,width=4in}
\vskip -.3cm
\caption{\sl The optical band-gap of CsCl decreases with increasing tensile 
stress.}
\vskip -1cm
\end{center}
\label{fig:uvplot}
\vskip -0.5cm
\end{figure}

The theoretical calculations also show a similar decreasing trend in band-gap 
with increasing lattice constant as our experimental results. However, fig~4 
does not indicate a large variation in band-gap with lattice constant as
indicated in fig~2. As stated earlier, our films were
colored indicating the existence of vacancies in our films. Hence we
repeated our calculations for band gap, however, in this case
with one chlorine atom absent per 27 unit cells. Fig~4 compares the two
results. A chlorine atom vacancy in 27 unit cell leads to a reduction of 
0.3eV in the material's band gap. This variation is nearly constant
throughout the range of increase in lattice constant studied. The 
scattering of our data points and large variation in our band gap (fig~2) 
might be explained by combined effect of increasing lattice constant and 
unsystematically variation in number of vacancies. We believe these 
characterizations can be of use for controlled fabrication
of optical devices or optical waveguides using CsCl.

\section{Conclusions}

The optical properties of thin films of CsCl are found to strongly depend on
the unit cell's lattice constant and the vacancies present in it. The
vacancies give rise to energy levels within the forbidden gap which results
in giving a color to the hitherto transparent film. Similarly, increasing
lattice size results in larger inter-atomic distances between neighboring
atoms that results in breaking of degeneracy of energy levels that in turn
affects the material's band gap.

\section*{Acknowledgment}

The help in completing the spectroscopic and diffraction analysis by 
Mr. Dinesh Rishi (USIC), Mr. Padmakshan and Mr. Rohtash (Department of Geology, 
Delhi University) is also acknowledged.
L.M. and E.A.A. acknowledge financial support from the Consejo Nacional de
Investigaciones Cient\'{\i}ficas y T\'ecnicas (CONICET), 
the Universidad Nacional de Entre R\'{\i}os (UNER), and the Agencia Nacional
de Promoci\'on Cient\'{\i}fica y Tecnol\'ogica (ANPCyT), Argentina.

\end{document}